\documentclass[pra,onecolumn]{revtex4}
\usepackage{amssymb}
\usepackage{amsmath}
\usepackage{graphicx}
\usepackage{dcolumn}
\usepackage[center]{subfigure}
\usepackage{array}
\usepackage{booktabs}
\usepackage{multirow}
\usepackage{color}

\begin{document}

\title{Semidiscrete vortex solitons}
\author{Xiaoxi Xu$^{1\S }$, Guanghao Ou$^{1\S }$, Zhaopin Chen$^{2}$, Bin Liu%
$^{1}$, Boris A. Malomed$^{2,3}$, Weicheng Chen$^{1}$, and Yongyao Li$^{1,2}$}
\email{yongyaoli@gmail.com}
\affiliation{$^{1}$School of Physics and Optoelectronic Engineering, Foshan University,
Foshan 528000, China \\
$^{2}$Department of Physical Electronics, School of Electrical Engineering,
Faculty of Engineering, and Center for Light-Matter Interaction, Tel Aviv
University, P.O.B. 39040, Tel Aviv, Israel\\
$^{3}$Instituto de Alta Investigaci\'{o}n, Universidad de Tarapac\'{a},
Casilla 7D, Arica, Chile\\
$\S $These two authors contribute equally to this paper}

\begin{abstract}
We demonstrate a possibility of creation of stable optical solitons
combining one continuous and one discrete coordinates, with embedded
vorticity, in an array of planar waveguides with intrinsic cubic-quintic
nonlinearity. The same system may be realized in terms of the spatiotemporal
light propagation in an array of tunnel-coupled optical fibers with the
cubic-quintic nonlinearity. In contrast with zero-vorticity states,
semidiscrete vortex solitons do not exist without the quintic term in the
nonlinearity. Two types of the solitons, \emph{viz.}, intersite- and
onsite-centered ones (IC and OC, respectively), with even and odd numbers $N$
of actually excited sites in the discrete direction, are identified. We
consider the modes carrying the embedded vorticity $S=1$ and $2$. In
accordance with their symmetry, the vortex solitons of the OC type exhibit
an intrinsic core, while the IC solitons with small $N$ may have a coreless
structure. Facilitating their creation in the experiment, the modes reported
in the present work may be much more compact states than their counterparts
considered in other systems, and they feature strong anisotropy. They can be
set in motion in the discrete direction, provided that the coupling constant
exceeds a certain minimum value. Collisions between moving vortex solitons
are considered too.\newline
\end{abstract}

\maketitle

\section{Introduction}

The creation of stable solitons in multidimensional geometry is a subject of
intensive ongoing research in nonlinear optics, Bose-Einstein condensates
(BECs), and other fields \cite{NaturePhysRev,Malomed2019}. A fundamental
problem which impedes straightforward making of such solitons is that the
ubiquitous cubic self-focusing nonlinearity gives rise to the critical and
supercritical collapse in the two- and three-dimensional (2D and 3D) space,
respectively \cite{Lberg,Fibich}, which destabilizes the solitons, on the
contrary to stable ones supported by the cubic nonlinearity in 1D.

Thus, stabilization of multidimensional solitons is the central problem in
this field. One possibility is the use of spatially periodic potentials,
induced by photonic crystals in optics \cite{Kivsharbook,Skoro,Kivshar2012}
and by optical lattices in BEC \cite{Baizakov,Jianke2003}. In the limit of
very deep periodic potentials, the nonlinear medium creates effectively
discrete solitons \cite%
{Malomed2001,Martin2004,Neshev2004,exper2,FLederer2008,in-book}. Recently,
another stabilization method was elaborated for two-component solitons,
based on the use of the linear spin-orbit coupling in binary BEC \cite%
{Sakaguchi2014,HPu,Malomed2018}, or its counterpart in the bimodal light
propagation in planar waveguides \cite{YVK2015}. STOP

The above-mentioned stabilization mechanisms for 2D and 3D solitons rely on
the use of linear effects. The stability may also be provided by replacement
of the cubic nonlinearity by other terms, which do not lead to the collapse.
In optics, stable 2D spatial solitons and spatiotemporal light bullets were
created in quadratic nonlinear crystals \cite{Torruellas1995,XLiu2000},
saturable \cite{Segev1994} and nonlocal \cite{Peccianti2002} nonlinear
media, and, finally, in a bulk material which features competing cubic and
quintic (CQ) focusing and defocusing interactions \cite{Edilson}. It is
relevant to mention that the CQ nonlinearity, with negligible corrections
from higher-order terms, was experimentally identified in optical materials
such as CS$_{2}$ (for power densities up to hundreds of GW/cm$^{2}$ \cite%
{CS2}) and colloidal suspensions of metallic nanoparticles \cite%
{colloid,colloid2}. In BEC, stable 2D matter-wave solitons were predicted in
dipolar BECs \cite{Pedri2005,Tikhonenkov2008} and microwave-coupled binary
condensates \cite{JliQin2016}, where the nonlinearity is cubic but
effectively nonlocal. Recently, experiments have revealed soliton-like
multidimensional matter-wave states, in the form of \emph{quantum droplets}
(QDs), filled by an incompressible ultradilute quantum fluid \cite{reviewQD}%
, in dipolar BECs \cite{Schmitt2016,Baillie2016,Chomaz2016}, and in binary
condensates with contact interactions \cite%
{Cabrera2018,Cabrera,Inguscio1,Inguscio2,hetero}. The latter experiments
followed the prediction of the stabilization of the QDs by the
Lee-Huang-Yang (LHY) correction to the mean-field interactions \cite{LHY},
reported in Refs. \cite{Petrov2015,Petrov2016}. The correction originates
from quantum fluctuations around the mean-field states.

A still more challenging issue is the creation of stable 2D and 3D bright
solitons with embedded vorticity, because such a state is subject to the
instability against splitting into fragments by azimuthal perturbations,
which is a more destructive factor (the one acting faster) than the collapse
\cite{Malomed2019}. Similar to their fundamental (zero-vorticity)
counterparts, bright vortex solitons can be stabilized by means of linear
effects, such as lattice potentials \cite{Baizakov,Jianke2003,Yiyin}, and
with the help of modified nonlinearity. In particular, effectively
two-dimensional stable discrete and lattice optical vortex solitons were
predicted \cite{Malomed2001} and created in the experiment \cite%
{Martin2004,Neshev2004,exper2}. Stable 2D solitons with intrinsic vorticity
(topological charge) $S=1$ in the uniform CQ medium were first predicted in
Ref. \cite{Quiroga}, which was then extended to $S\geq 2$ \cite%
{Pego,Davydova2004}, and later to nonlinear lattices of the CQ type \cite%
{Gao2018}. The existence of stable 3D solitons with $S=1$ in the same model
was reported in Ref. \cite{Mihalache2002}. In binary BECs, stable 2D vortex
solitons with topological charges up to $S=5$ have been predicted in the
above-mentioned microwave-coupled condensates, as well as in QDs with
embedded vorticity \cite{yongyao20182DVQD,Zeda}. Stable 3D QDs filled by the
\textquotedblleft swirling" condensate with $S=1$ and $2$ were also revealed
by the analysis \cite{Barcelona}.

A relevant extension of the analysis of topological solitons aims to predict
stable vortex modes with an anisotropic shape. In optics, stable 2D discrete
vortex states were predicted in 2D anisotropic discrete lattices \cite%
{PGK2005,Guihua2015}. In BEC, vortex solitons were predicted in anisotropic
lattice potentials \cite{Baizakov2}, as well as in one component of the
spin-orbit-coupled binary BEC with the anisotropic dipole-dipole interaction
between atoms \cite{Xunda2016,gapSOCddi,bingjin2017}. Very recently,
creation of a novel form of 2D anisotropic vortices, \textit{viz}., \textit{%
semidiscrete} QDs with imprinted vorticity, was elaborated in a system with
one continuous and one discrete coordinates, realized as a binary condensate
loaded in an array of parallel tunnel-coupled quasi-1D traps \cite{semiVQD}.
In that model, the intrinsic nonlinearity of the traps is represented by the
combination of self-attractive quadratic LHY and repulsive cubic mean-field
terms. A nontrivial peculiarity of the setting is identification of the
vorticity, which is defined in terms of the global phase pattern carried by
the semidiscrete state.

\begin{figure}[h]
{\includegraphics[width=0.4\columnwidth]{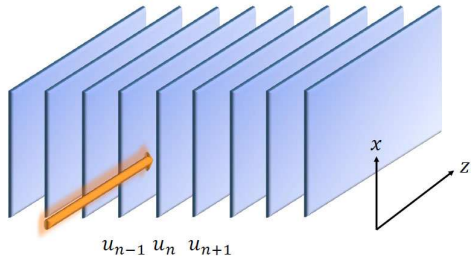}}
\caption{The setting based on the array of planar optical waveguides (blue
slabs), separated by gray isolating layers, with the continuous transverse
coordinate, $x$, and the discrete one, $n$. As shown by the arrow, light is
coupled into the array along the $z$ direction.}
\label{Model}
\end{figure}

Semidiscrete vortices were not yet considered in nonlinear optics, which is
the subject of the present work. Because arrays of coupled nonlinear
waveguides are ubiquitous systems in optics \cite{FLederer2008,in-book}, the
creation of nontrivial self-trapped states in the arrays is a very relevant
objective. The present work aims to predict stable semidiscrete optical
vortex solitons in a effectively 2D setting built as an array of parallel
tunnel-coupled planar waveguides with the intrinsic CQ nonlinearity, as
sketched in Fig. \ref{Model}. This setting may be naturally considered as a
semidiscrete one. The respective model is somewhat similar to the
above-mentioned one developed for the array of BEC traps with the intrinsic
quadratic-cubic nonlinearity, which also supports stable vortex solitons
\cite{semiVQD}. However, results concerning the shape and stability of
semidiscrete vortex solitons in the present system are essentially
different. In particular, they may feature strong anisotropy, unlike
quasi-isotropic modes reported in Ref. \cite{semiVQD}, and they may be built
with much tighter shapes.

The rest of the paper is structured as follows. The model is introduced in
Sec. II, including two possible realizations in optics, \textit{viz.}, in
terms of the light propagation in the spatial domain, as outlined in Fig. %
\ref{Model}, and in the spatiotemporal one, realized as an array of
tunnel-coupled fiber-like waveguides. Systematic results for the existence
and stability of vortex solitons with $S=1$ and $S=2$ are summarized in Sec.
III. The paper is concluded by Sec. IV.

\section{The model}

The propagation of light in the waveguiding array displayed in Fig. \ref%
{Model} is modeled, in the paraxial approximation, by coupled nonlinear Schr%
\"{o}dinger equations (NLSEs) \cite{Aceves,Panoiu2006} with the CQ terms
\cite{Suba2007}:
\begin{eqnarray}
&&2ik_{0}{\frac{\partial }{\partial Z}}A_{n}=-{\frac{\partial ^{2}}{\partial
X^{2}}}A_{n}-\kappa \left( A_{n+1}-2A_{n}+A_{n-1}\right)   \notag \\
&&-2k_{0}^{2}{\frac{{\large n}_{2}}{{\large n}_{0}}}%
|A_{n}|^{2}A_{n}+2k_{0}^{2}{\frac{{\large n}_{4}}{{\large n}_{0}}}%
|A_{n}|^{4}A_{n}-i\zeta A_{n}.  \label{fullNLSE}
\end{eqnarray}%
where $A_{n}$ is the envelope of the electromagnetic field in the $n$-th
guiding core, with carrier wavenumber $k_{0}=2{\large n}_{0}\pi /\lambda $, $%
\kappa $ is the strength of the tunnel coupling between adjacent cores, $%
{\large n}_{0}$, ${\large n}_{2}$ and ${\large n}_{4}$ represent,
respectively, the linear, third- and fifth-order refractive indices of the
medium, and $\zeta $ is the loss coefficient. Equation (\ref{fullNLSE}) does
not include multiphoton absorption (nonlinear losses), as we do not consider
media with resonant interactions or ionization of the optical material,
which would induce such losses \cite{multi}. The coupling constant is
determined by the difference of the refractive index between the guiding
cores and dielectric material which separates them, the carrier wavelength,
and the thickness of the separating layer, exponentially decaying with the
increase of the latter. In practical terms, experimentally relevant values
of the thickness are tantamount to a few wavelengths (which corresponds to
several microns). This choice leads to the propagation length, for which the
coupling effects become essential, measured in several millimeters \cite%
{FLederer2008}. As concerns the losses, in available optical materials they
take values \symbol{126}dB/m or still smaller, which makes the losses
negligible for experimentally available propagation lengths $\lesssim 10$
cm. For this reason, the dissipative term is neglected in the following
consideration.

The semidiscrete systems adequately modeled by Eq. (\ref{fullNLSE}) with the
cubic-only nonlinearity ($n_{4}=0$), were first introduced, and soliton-like
states in them were investigated, in Refs. \cite%
{Aceves,Aceves2,Aceves3,Aceves4,Aceves5,Blit}, in terms of the
spatiotemporal propagation of light in arrays of optical fibers. On the
other hand, the 2D discrete model with the CQ onsite nonlinearity, and
discrete soliton modes in them, were considered in Ref. \cite{CQdiscr2}. A
2D quasi-discrete model, combining a deep checkerboard potential and the CQ
nonlinearity, was also addressed, along with its vortex-soliton modes, in
Ref. \cite{checker}.

By applying rescaling,
\begin{eqnarray}
&&U_{n}=\sqrt{{\large n}_{4}/{\large n}_{2}}A_{n},\quad C=\left( {\large n}%
_{0}{\large n}_{4}/2k_{0}^{2}{\large n}_{2}^{2}\right) \kappa   \notag \\
&&z=\left( k_{0}{\large n}_{2}^{2}/{\large n}_{0}{\large n}_{4}\right)
Z,\quad x=Xk_{0}{\large n}_{2}/\sqrt{{\large n}_{0}{\large n}_{4}},
\label{rescale}
\end{eqnarray}%
Eq. (\ref{fullNLSE}) is cast in the normalized form,
\begin{eqnarray}
&&i\partial _{z}U_{n}=-{\frac{1}{2}}\partial _{xx}U_{n}-{\frac{C}{2}}\left(
U_{n+1}-2U_{n}+U_{n-1}\right)   \notag \\
&&-|U_{n}|^{2}U_{n}+|U_{n}|^{4}U_{n},  \label{scalNLSE}
\end{eqnarray}%
with the effective intersite coupling constant, $C$. The total power of the
field in the scaled form, which is a dynamical invariant of Eq. (\ref%
{scalNLSE}), is defined as
\begin{equation}
P=\sum_{n=-\infty }^{+\infty }\int_{-\infty }^{+\infty }|U_{n}(x)|^{2}dx.
\label{Power}
\end{equation}%
The model also conserves the Hamiltonian,%
\begin{equation}
H=\frac{1}{2}\sum_{n=-\infty }^{+\infty }\int_{-\infty }^{+\infty }\left[
|\partial _{x}U_{n}|^{2}+C\left\vert U_{n}-U_{n-1}\right\vert
^{2}-|U_{n}|^{4}+\frac{2}{3}|U_{n}|^{6}\right] dx.  \label{H}
\end{equation}

To estimate physical parameters of the model, we assume that, for instance,
the cubic-quintic material is CS$_{2}$, which was used for the creation of
stable fundamental 2D solitons in Ref. \cite{Edilson}. At the carrier
wavelength $\lambda =800$ nm, its parameters are $n_{0}=1.61$, ${\large n}%
_{2}=3.1\times 10^{-19}$ m$^{2}$/W, and $n_{4}=5.2\times 10^{-35}$ m$^{4}$/W$%
^{2}$ \cite{Couris2003,Dkong2009}. Then, relations between scaled units in
Eq. (\ref{scalNLSE}) and physical ones are estimated, by means of Eq. (\ref%
{rescale}), as follows: $x=1\Longleftrightarrow 0.235$ $\mathrm{\mu }$m, $%
z=1\Longleftrightarrow 0.07$ mm, and $P=1\Longleftrightarrow 70$ kW, if the
thickness of the single planar waveguide is $0.5$ $\mathrm{\mu }$m. As
mentioned above, the losses in the medium may be neglected for
experimentally relevant propagation distances \cite{Edilson}. With respect
to these estimates, the characteristic transverse size of self-trapped modes
considered below, $\Delta x\sim 20-40$, corresponds to the physical width $%
\sim 5-10$ $\mathrm{\mu }$m, and the characteristic propagation distance,
which is $z\sim 500-3000$, amounts to $3.5\sim 20$ cm. These values are
realistic for feasible experiments, see Refs. \cite%
{Martin2004,Neshev2004,exper2,FLederer2008}. Note also that $\Delta x$ is
sufficiently large in comparison to $\lambda $, which justifies the use of
the paraxial propagation equation (\ref{scalNLSE}).

The same model applies to the temporal-domain light propagation, with
carrier group velocity $V_{\mathrm{gr}}$, in an array of tunnel-coupled
optical fibers with the intrinsic CQ nonlinearity, if coordinate $x$ in Eq. (%
\ref{scalNLSE}) is replaced by the temporal coordinate, $\tau =t-z/V_{%
\mathrm{gr}}$. In that case, the solitons represent semidiscrete
\textquotedblleft light bullets", and Eq. (\ref{Power}) defines the total
energy of the spatiotemporal optical signal.

\section{Results}

\subsection{Semidiscrete vortex solitons with topological charge $S=1$}

\begin{figure}[tph]
{\includegraphics[width=0.5\columnwidth]{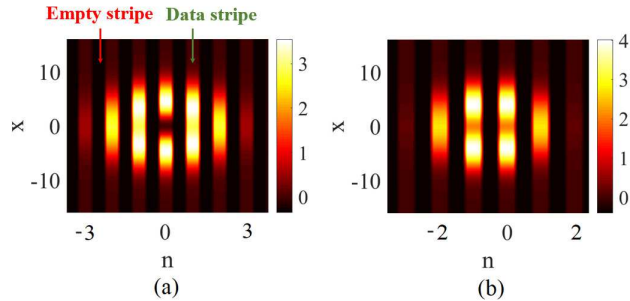}}
\caption{Typical examples of the inputs used for the generation of vortex
solitons of the OC (a) and IC (b) types, taken as per Eqs. (\protect\ref%
{initial}) and, respectively, (\protect\ref{OC}) or (\protect\ref{IC}).
Input (a) with parameters $(A,\protect\alpha ,P)=(0.85,0.0015,50)$ produces
the OC soliton shown in Fig. \protect\ref{ICOCexample}(f). Input (b) with
parameters $(A,\protect\alpha ,P)=(0.09,0.0015,40)$ produces the IC soliton
shown in Fig. \protect\ref{ICOCexample}(b).}
\label{Initialguess}
\end{figure}

\begin{figure}[tph]
{\includegraphics[width=1.0\columnwidth]{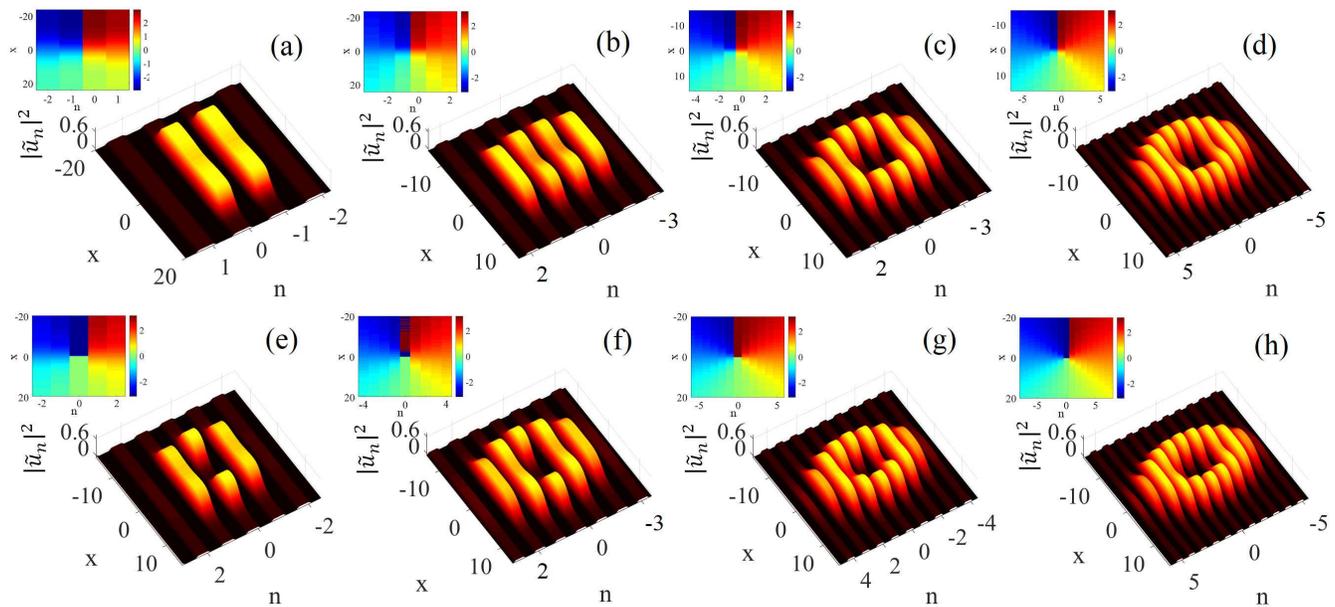}}
\caption{(a-d) Typical examples of the intensity pattern of stable IC vortex
solitons with $S=1$, which correspond to points \textquotedblleft A-D" in
the stability areas in Fig. \protect\ref{stablearea}(a). Their parameters
are $(P,C,N,N_{\mathrm{core}})=(39,0.01,2,0)$ (a), ($40,0.05,4,0$) (b), ($%
55,0.13,6,2$) (c), and ($70,0.21,8,2$) (d). (e-f) Typical examples of the
intensity pattern for stable OC vortex solitons with $S=1$, which correspond
to points \textquotedblleft E-H" in the stability areas in Fig. \protect\ref%
{stablearea}(b). The corresponding parameters are $(P,C,N,N_{\mathrm{core}%
})=(32,0.035,3,1)$ (e), ($50,0.1,5,1$) (f), ($65,0.18,7,3$) (g), and ($%
80,0.27,9,3$) (h). Insets display the respective phase patterns, which
identify the soliton's vorticity, $S=1$.}
\label{ICOCexample}
\end{figure}

Stationary solutions to Eq. (\ref{scalNLSE}) with propagation constant $%
\beta $ are looked for as
\begin{equation}
U_{n}(x,z)=u_{n}(x)e^{i\beta z}.  \label{beta}
\end{equation}%
Stationary semidiscrete vortex solitons with topological charge $S$, of the
onsite-centered (OC) and intersite-centered (IC) types, with the pivot
located, respectively, at a lattice site or between two sites, were produced
by means of the power-conserving squared-operator numerical method \cite%
{PCSOM}, initiated by inputs
\begin{equation}
\phi _{0}=AR^{S}\exp (iS\Theta -\alpha R^{2}),  \label{initial}
\end{equation}%
where $A$ and $\alpha $ are positive real constants. For the OC and IC
solitons, we set in Eq. (\ref{initial})
\begin{equation}
R_{\mathrm{OC}}^{2}=x^{2}+(n/\sqrt{C})^{2},\Theta _{\mathrm{OC}}=\arctan (n/%
\sqrt{C}x),  \label{OC}
\end{equation}%
\begin{equation}
R_{\mathrm{IC}}^{2}=x^{2}+[(n+{1/2})/\sqrt{C}]^{2},\Theta _{\mathrm{IC}%
}=\arctan \left[ \left( n+1/2\right) /\sqrt{C}x\right] ,  \label{IC}
\end{equation}%
respectively.

Examples of the inputs used for the generation of the OC and IC solitons are
displayed in Fig. \ref{Initialguess}. To make the pictures clearer, in this
and other figures values of local intensities, $\left\vert
U_{n}(x)\right\vert ^{2}$, are shown, at discrete values of $n$, by means of
finite-width stripes separated by empty ones [\textquotedblleft data" and
\textquotedblleft empty" stripes in Fig. \ref{Initialguess}(a)].

Then, stability of the stationary solutions was verified by direct
simulations of Eq. (\ref{scalNLSE}) for propagation distance $z=1000$. The
soliton is stable if its intensity profile stays unchanged throughout the
simulation. According to the above-mentioned estimate of parameters for CS$%
_{2}$, $z=1000$ corresponds to $Z=7$ cm, which is sufficiently long for
observing a stable spatial soliton in the experiment. In the numerical
computations, control parameters are total power $P$ (\ref{Power}) and
coupling constant $C$ in Eq. (\ref{scalNLSE}). In terms of vortex modes
displayed below in Fig. \ref{ICOCexample}, $z=1000$ corresponds to $\gtrsim
10$ characteristic Rayleigh (diffraction) lengths with respect to the
continuous and discrete coordinates, hence this distance is sufficient to
make conclusions concerning the stability of the modes.

\begin{figure}[h]
{\includegraphics[width=0.9\columnwidth]{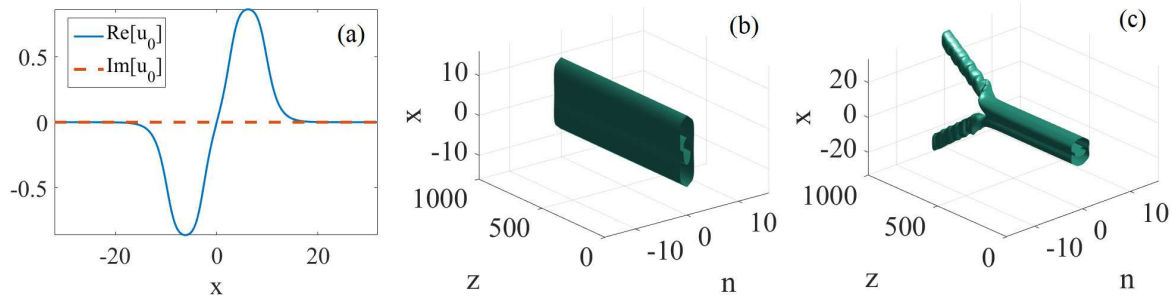}}
\caption{(a) Real and imaginary parts of the stationary wave function of the
OC soliton from Fig. \protect\ref{ICOCexample}(e) in the central ($n=0$)
waveguide. As explained in the text, the stationary wave field at the center
of OC vortex solitons with $S=1$ is real and spatially odd. (b) Direct
simulations of the perturbed evolution of a stable OC soliton with $%
(P,C,N,N_{\mathrm{core}})=(32,0.035,3,1)$. (c) Direct simulations for an
unstable OC soliton with $(P,C,N,N_{\mathrm{core}})=(33,0.07,3,1)$. }
\label{stableS1}
\end{figure}

Typical examples of the vortex solitons of the OC and IC types are displayed
in Fig. \ref{ICOCexample} [the sign of the vorticity, $S=1$, is determined
by the comparison of phase patterns in the figure with the standard
expression, $u\sim \exp \left( iS\Theta \right) $, in the continuum
coordinate plane with angular coordinate $\Theta $]. The solitons are
characterized by the number of effectively excited sites, $N$ (individual
waveguides in which the field takes non-negligible values). Another
important characteristic is number $N_{\mathrm{core}}$ of sites in the \emph{%
core} of the semidiscrete vortex, i.e., waveguides in which stationary
fields $u_{n}(x)$ [see Eq. (\ref{beta})] cross zero (vanish) at $x=0$, thus
having opposite signs at $x>0$ and $x<0$, while zero crossing is absent in
waveguides which do not belong to the core. Although fundamental
semidiscrete solitons, with zero vorticity, are also characterized by finite
$N$ \cite{Aceves,Aceves2,Aceves3,Aceves4,Aceves5,Blit}, only vortices may
feature the core. Examples displayed in Fig. \ref{ICOCexample} exhibit
values of the excited and in-core sites $2\leq N\leq 9$ and $0\leq N_{%
\mathrm{core}}\leq 3$, with odd and even numbers $N$ and $N_{\mathrm{core}}$
pertaining, severally, to the vortex modes of the OC and IC types. Note that
the states of the former type always have $N_{\mathrm{core}}\geq 1$, and,
due to their intrinsic symmetry, any vortex soliton of the OC type with odd $%
S$ has a real odd modal function $u_{0}(x)$ in the central waveguide ($n=0$%
), see Fig. \ref{stableS1}(a) below. On the other hand, the IC vortex states
may exist both with $N_{\mathrm{core}}=0$ and $N_{\mathrm{core}}\geq 2$, see
panels (a,b) and (c,d) in Fig. \ref{ICOCexample}. In the case of $N_{\mathrm{%
core}}=0$, the vorticity is accounted for by opposite signs of fields $%
u_{0}(x)$ and $u_{1}(x)$ in Eq. (\ref{beta}), with the \textquotedblleft
virtual pivot" of the vortex set between $n=0$ and $n=1$. %
It is also worthy
to note that the semidiscrete vortex solitons may be strongly anisotropic,
elongated in the continuous direction, in the case of small $C$, see, e.g.,
Figs. \ref{ICOCexample}(a,e). On the other hand, Fig. \ref{Chara}, displayed
below, demonstrates that stable vortices slightly elongated in the
discrete direction, exist too, corresponding to $\varepsilon<1$ in Fig. \ref{Chara}.

\begin{figure}[t]
{\includegraphics[width=0.6\columnwidth]{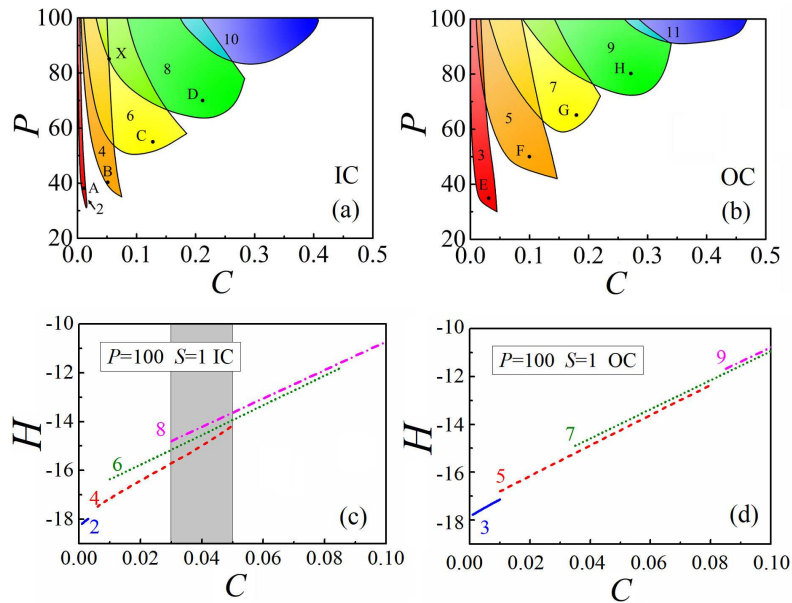}}
\caption{(Color online) Stability areas for semidiscrete vortex solitons of
the IC (a) and OC (b) types, in the $(P,C)$ plane (the total power and
intersite coupling constant). Vortex-soliton species are characterized by
the number of excited sites, $N$, and the size of the vortex's core, $N_{%
\mathrm{core}}$, in the discrete direction. These numbers, which are even in
(a) and odd in (b), label particular colored stability regions. (c,d) $H$ as
a function of $C$ at $P=100$ for stable vortex solitons of the IC and OC
type with different values of $N$ and $N_{\mathrm{core}}$. The tristability
for a given value of $P$ occurs in the gray stripe in panel (c), where,
obviously, the vortex mode with the smallest values of $N$ and $N_{\mathrm{%
core}}$ realizes the system's ground state.}
\label{stablearea}
\end{figure}

Stability areas for the OC and IC species of semidiscrete vortex solitons,
which are characterized by numbers $N$, are displayed in the $(P,C)$ plane
in Figs. \ref{stablearea}(a,b). A remarkable feature of the stability chart
is \emph{bistability}, allowing coexistence of the vortices with equal
powers $P$ and different numbers of sites, and even \emph{tristability} --
in particular, the overlap of the stability regions for the vortices of the
IC type with $N=4,6,8$, at $P>83$ [see point X in Fig. \ref{stablearea}(a)].
In fact, the multistability area is even broader, as Fig. \ref{stablearea}%
(b) demonstrates that, in addition to these three IC modes, two stable
vortices of the OC type exist too, with $N=5$ and $7$, at the same values of
$C$ and $P$. In the limit of $P\rightarrow \infty $, which corresponds to
the 2D\ continuum space, the multistability agrees with the known fact that
2D vortex solitons of the nonlinear Schr\"{o}dinger equation with CQ
nonlinearities become stable as the solitons expand to accommodate
indefinitely growing values of the norm \cite{Pego}. The calculation of
values of the Hamiltonian, according to Eq. (\ref{H}), demonstrates that the
minimum of the Hamiltonian, i.e., the ground state of the system, is
realized, in the case of bi- or tristability, by the mode with the smallest
number of excited discrete sites. An example of this calculation is shown in
Fig. \ref{stablearea}(c,d). Typically, the tristability area for the IC
solitons with $N=4$, $6$ and $8$ is shaded in Fig. \ref{stablearea}(c).
Recently,  multistability has also been
found in 1D quantum droplets trapped in a lattice potential \cite{liangwei}.

Outside of the stability area but close to its border, unstable vortex
solitons can also be found; in direct simulations, they split in fragments.
Typical examples of the evolution of stable and unstable vortex solitons are
displayed in Fig. \ref{stableS1}. Far from the stability boundary of the
vortices, only fundamental (zero-vorticity) solitons are produced by the
numerical calculations.

\begin{figure}[h]
{\includegraphics[width=0.6\columnwidth]{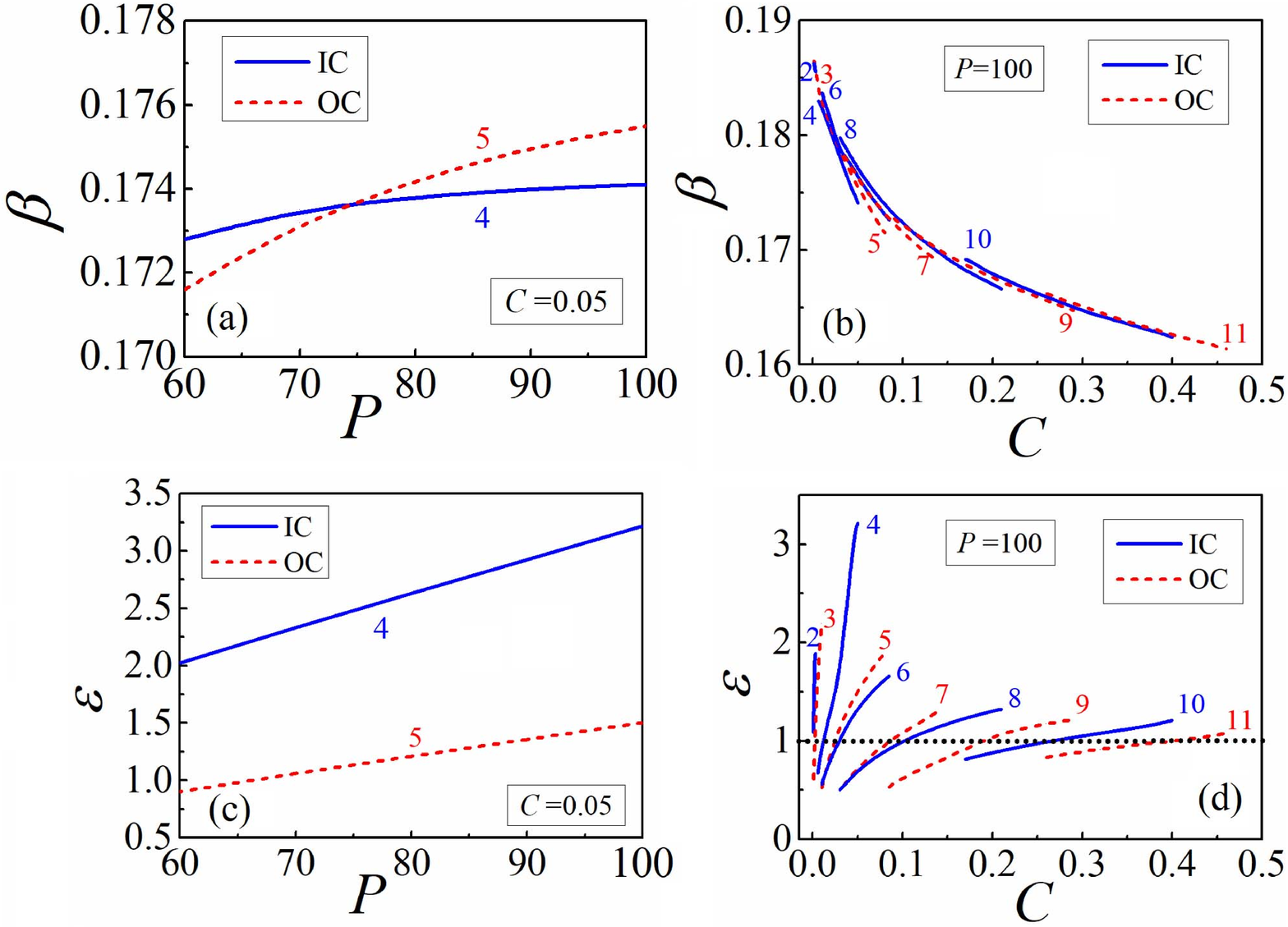}}
\caption{(a) Propagation constant $\protect\beta $ [see Eq. (\protect\ref%
{beta})] vs. the total power, $P$, for families of stable solitons of the IC
and OC types, which contain $N=4$ or $5$ excited sites in the discrete
direction, respectively, for $C=0.05$. The corresponding values of the
core's size are $N_{\mathrm{core}}=0$ and $1$. (b) $\protect\beta $ vs. the
coupling constant, $C$, for the families of stable solitons with different
numbers of $N$ and $N_{\mathrm{core}}$. Here, we fix $P=100$. (c) The
anisotropy measure $\protect\varepsilon $, defined as per Eq. (\protect\ref%
{aniso}) for the vortex solitons, with $\left( N,N_{\mathrm{core}}\right)
=(4,0)$ and $\left( N,N_{\mathrm{core}}\right) =(5,1)$, vs. $P$, for $C=0.05$.
(d) $\protect\varepsilon $ for the stable vortex states, with different
values of $N$ and $N_{\mathrm{core}}$, as a function of $C$, for fixed $%
P=100 $. Extensions of the displayed branches (which are not shown in the
figure) are formed by unstable solitons.}
\label{Chara}
\end{figure}

Note that the vortex solitons of the IC and OC types with only two and three
excited sites (and, respectively, $N_{\mathrm{core}}=0$ and $N_{\mathrm{core}%
}=1$) realize two types of the solitons with the minimum size in the
discrete direction. Such tightly localized semidiscrete modes were not
reported in the previously studied semidiscrete system \cite{semiVQD}. In
fully discrete 2D models, the possibility that the smallest vortex soliton
includes four sites was theoretically predicted \cite{Malomed2001} and
experimentally demonstrated \cite{Neshev2004,exper2}.

Figure \ref{Chara}(a) displays the propagation constant, $\beta $, for two
types of compact solitons [IC and OC\ with $(N,N_{\mathrm{core}})=(4,0)$ and
$(N,N_{\mathrm{core}})=(5,1)$, respectively] as a function of the total
power, $P$, for fixed $C$. These results indicate that the $\beta (P)$
curves satisfy the Vakhitov-Kolokolov criterion, $d\beta /dP>0$, which is a
well-known necessary stability condition \cite{Vakh,Lberg}. Figure \ref%
{Chara}(b) shows the $\beta (C)$ dependence for stable vortex solitons with
different numbers of excited sites $N$ and fixed $P$, showing that $\beta $
gradually decreases with $C$.

To characterize anisotropy of the vortex soliton, we define parameter
\begin{equation}
\varepsilon =\sqrt{C}{\frac{L_{x}}{L_{n}}},  \label{aniso}
\end{equation}
\begin{equation}
L_{x}\equiv {\frac{\left( \int_{-\infty }^{+\infty }|u_{0}(x)|^{2}dx\right)
^{2}}{\int_{-\infty }^{+\infty }|u_{0}(x)|^{4}dx}},\quad L_{n}\equiv {\frac{%
\left( \sum_{n}|u_{n}(x=0)|^{2}\right) ^{2}}{\sum_{n}|u_{n}(x=0)|^{4}}}.
\label{LxLn}
\end{equation}%
This definition is adopted from Ref. \cite{semiVQD}, where $\varepsilon =1$
implied that solitons were effectively isotropic modes, while $\varepsilon
>1 $ and $\varepsilon <1$ indicated that their shape was anisotropic,
namely, elongated in the continuous or discrete direction, respectively.
Figure \ref{Chara}(c) displays $\varepsilon $ as a function of $P$ for the
compact vortex solitons with $(N,N_{\mathrm{core}})=(4,0)$ and $(N,N_{%
\mathrm{core}})=(5,1)$. This figure indicates that $\varepsilon $ grows,
approximately, as a linear function of $P$. Figure \ref{Chara}(d) displays
the $\varepsilon (C)$ curves for vortex solitons with different values of $N$
and $N_{\mathrm{core}}$, and fixed $P$. The figure shows that the slope of $%
\varepsilon (C)$ decreases with the increase of $N$. Naturally, values of $%
\varepsilon (C)$ get closer to the isotropy point, $\varepsilon =1$, for
larger $N$, as this case corresponds to the quasi-continuum limit of the
system in the discrete direction, with the vortex soliton getting less
compact. Moreover, passage of $\varepsilon (P)$ and $\varepsilon (C)$
through the level of $\varepsilon =1$ implies that the vortex solitons may
be tuned to the isotropic shape by selecting specific values of $P$ or $C$.

\subsection{Effect of the relative strengths of the cubic and quintic
nonlinearities}

\begin{figure}[h]
{\includegraphics[width=0.6\columnwidth]{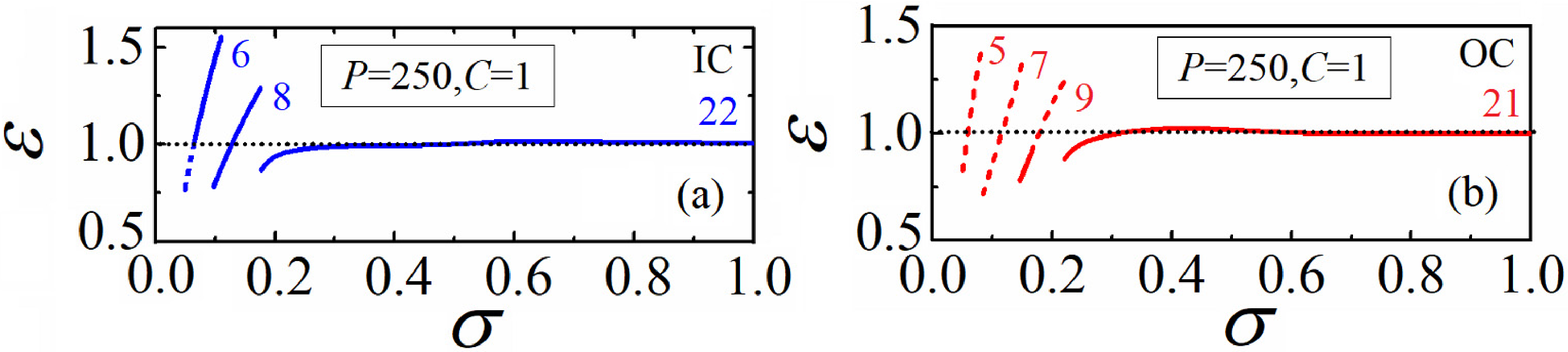}}
\caption{Anisotropy parameter $\protect\varepsilon $ as a function of the
relative strength of the quintic nonlinearity, $\protect\sigma $, for
semidiscrete vortex solitons with $S=1$, of the IC (a) and OC (b) types, as
produced by Eq. (\protect\ref{sigma}), with fixed total power $P=250$. Solid
and dashed curves represent stable and unstable solutions, respectively.
Numbers attached to different segments denote the number $N$ of excited
sites in the respective semidiscrete modes [even values $8<N<22$ in (a), and
odd ones $9<N<21$ in (b) show a gradual transition between the presented
branches]. Long branches close to $\protect\varepsilon =1$ correspond to
broad quasi-continuum modes with large values of $N$ [in particular, with $%
N=22$ and $21$, as indicated in panels (a) and (b)].}
\label{epsilonOCICsigma}
\end{figure}

A natural question is how the competing self-focusing cubic and defocusing
quintic onsite terms in Eq. (\ref{scalNLSE}) affect the existence and
stability of semidiscrete vortex solitons. To address this issue, we here
rescale the equation, to make the intersite coupling constant equal to $1$
and admit the presence of a free coefficient in front of the quintic term as
a free parameter:
\begin{equation}
\tilde{U}_{n}=\sqrt{C}U_{n},\tilde{z}=z/C,\tilde{x}=x/\sqrt{C},
\label{tilde}
\end{equation}%
hence the rescaled total norm is $\tilde{P}=\sqrt{C}P$. The substitution of
rescaling (\ref{tilde}) in Eq. (\ref{scalNLSE}) casts it in the following
form, where we omit the tilde, and use symbol $\sigma \equiv C$, to stress
that $C$ appears here as a newly defined parameter controlling the relative
strength of the quintic term:
\begin{eqnarray}
&&i\partial _{\tilde{z}}\tilde{U}_{n}=-{\frac{1}{2}}\partial _{\tilde{x}%
\tilde{x}}\tilde{U}_{n}-{\frac{1}{2}}\left( \tilde{U}_{n+1}-2\tilde{U}_{n}+%
\tilde{U}_{n-1}\right)  \notag \\
&&-|\tilde{U}_{n}|^{2}\tilde{U}_{n}+\sigma |\tilde{U}_{n}|^{4}\tilde{U}_{n}.
\label{sigma}
\end{eqnarray}

To characterize effects of the competition between the cubic and quintic
nonlinearities, Fig. \ref{epsilonOCICsigma} displays anisotropy parameter $%
\varepsilon $ [see Eq. (\ref{aniso})] as a function of $\sigma $ for the IC
and OC vortex solitons, as obtained from the numerical solution of Eq. (\ref%
{sigma}) with a fixed total power, $P=250$. The figure shows that, at $%
\sigma >0.2$, $\varepsilon $ stays very close to $\varepsilon =1$, which
indicates that the semidiscrete vortex solitons keep an effectively
isotropic profile (hence, they have a broad quasi-continuum shape) in this
case. At $\sigma <0.2$, the vortex solitons feature a compact shape,
adequately characterized by the number of excited sites, $N$. The segments,
which are characterized by the number of excited sites $5\leq N\leq 9$,
terminate in Figs. \ref{epsilonOCICsigma}(a,b) at their top and bottom
points because solutions could not be found above or below them. At still
smaller values of $\sigma $ the vortex solitons are unstable, and no vortex
modes were found at $\sigma \rightarrow 0$, when the CQ nonlinearity turn
into the cubic form. Thus, the inclusion of the self-defocusing quintic term
\emph{is necessary} for the existence and stability of the semidiscrete
vortex solitons.

\subsection{Vortex solitons with $S=2$}

\begin{figure}[!htp]
{\includegraphics[width=1.0\columnwidth]{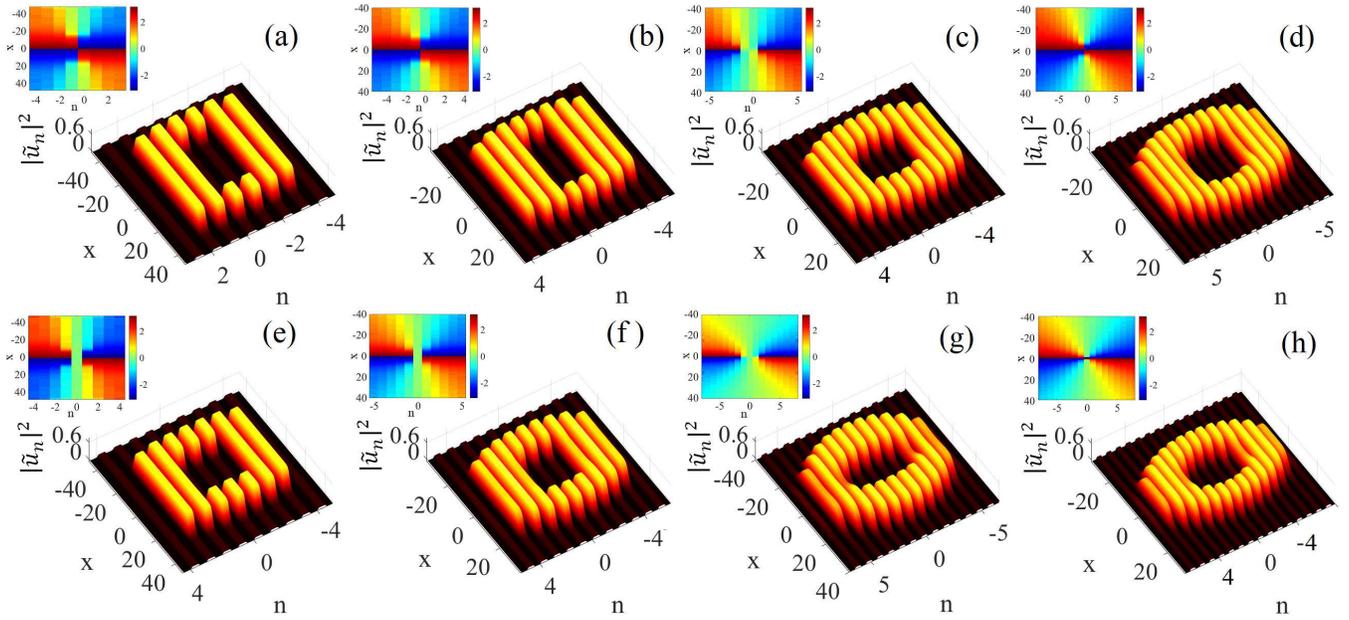}}
\caption{(a-d) Typical examples of the intensity pattern for stable IC
vortex solitons with $S=2$, for parameters $\left( P,C,N,N_{\mathrm{core}%
}\right) =(250,0.01,6,2)$ (a), ($250,0.03,8,2$) (b), ($250,0.04,10,4$) (c),
and ($250,0.1,12,4$) (d). (e-f) The same for stable vortex solitons with $%
S=2 $ of the OC type, with $\left( P,C,N,N_{\mathrm{core}}\right)
=(250,0.01,7,3) $ (e), ($250,0.04,9,3$) (f), ($250,0.1,11,5$) (g), and ($%
250,0.12,13,5$) (h). Insets display the respective phase patterns, which
identify the vorticity, $S=2$. }
\label{ICOCexampleS2}
\end{figure}

\begin{figure}[!h]
{\includegraphics[width=0.6\columnwidth]{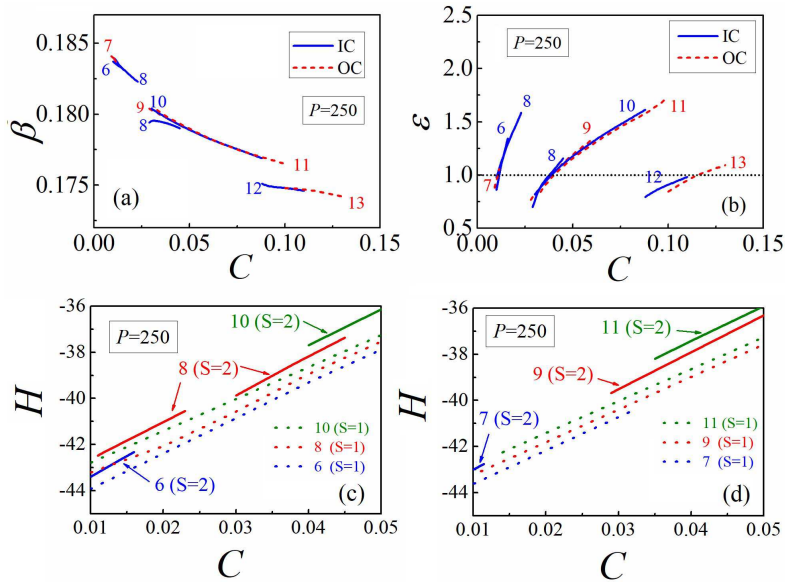}}
\caption{(a) Dependences $\protect\beta (C)$ for stable vortex solitons with
$S=2$, fixed total power, $P=250$, and different numbers $N$ of excited
sites (extensions of the branches, which are not shown here, are formed by
unstable solitons). (b) The same as in (a) but showing $\protect\varepsilon %
(C)$ dependences for stable vortex solitons with $S=2$. (c,d) The comparison
of values of the Hamiltonian of the vortex solitons with $S=1$ and $2$
(dotted and solid segments, respectively), and equal total powers, $P=250$,
of the IC (c) and OC (d) types.}
\label{charS2}
\end{figure}

A challenging issue is whether stable vortex solitons can be found for the
double topological charge, $S=2$. Numerical results demonstrate that such
solitons indeed exist -- naturally, with larger powers than their
counterparts with $S=1$. They can be found, at least, at $P>200$. Here, we
fix $P=250$ and consider characteristics of the vortex solitons with $S=2$,
varying intersite coupling $C$. The minimum numbers of sites necessary for
constructing the double-vortex modes of the IC and OC types are $N=6$ (with $%
N_{\mathrm{core}}=2$) and $N=7$ (with $N_{\mathrm{core}}=3$), respectively.
Typical examples for $6\leq N\leq 13$ and $2\leq N_{\mathrm{core}}\leq 5$
(even and odd $N$ and $N_{\mathrm{core}}$ for the solitons of the IC and OC
types, respectively) are displayed in Fig. \ref{ICOCexampleS2}, and
dependences $\beta (C)$ and $\varepsilon (C)$ for stable vortex solitons
with $S=2$ and different values of $N$ are displayed in Fig. \ref{charS2}.

\begin{figure}[!h]
{\includegraphics[width=0.6\columnwidth]{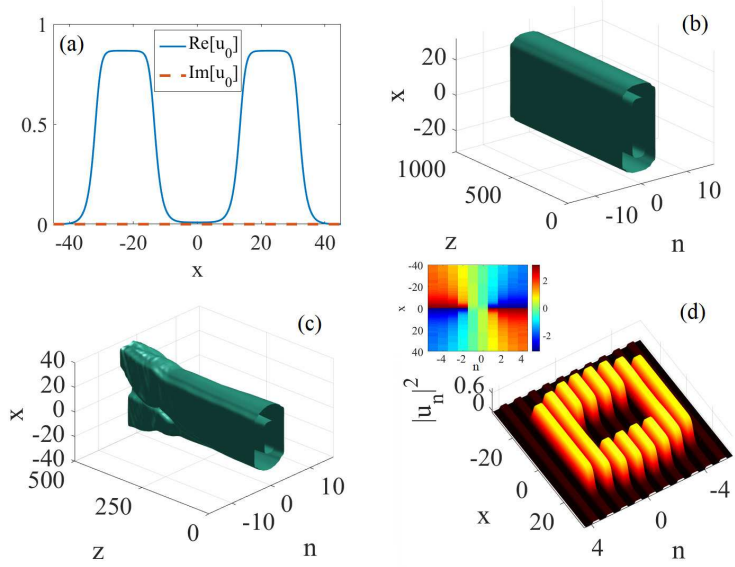}}
\caption{(a) Real and imaginary parts of the OC soliton from Fig. \protect
\ref{ICOCexampleS2}(e) in the central waveguide, $n=0$. (b) Direct
simulations of the perturbed evolution of a stable vortex soliton of the IC
type, with $S=2$ and $(P,C,N,N_{\mathrm{core}})=(250,0.03,8,2)$. (c) The
perturbed evolution of an unstable soliton of the same type, with $(P,C,N,N_{%
\mathrm{core}})=(250,0.25,8,2)$. (d) A typical example of the intensity
pattern for stable IC vortex solitons with double vorticity, $S=2$, the
parameters being $\left( P,C,N,N_{\mathrm{core}}\right) =(250,0.018,8,4)$.
The subplot is the respective phase pattern, which identifies the vorticity,
$S=2$. In comparison to its counterpart displayed in Fig. \protect\ref%
{ICOCexampleS2}(b), this soliton represents the branch corresponding to
smaller values of $C$, see further details in the text.}
\label{unstableS2}
\end{figure}

The intrinsic symmetry of the vortex solitons of the OC type, with even
vorticity $S\geq 2$, predicts that its modal wave function in the central
waveguide ($n=0$), $u_{0}(x)$, is a real even function vanishing at $x=0$.
Indeed, an example displayed in Fig. \ref{unstableS2}(a) clearly
corroborates the prediction. This feature may be compared to its counterpart
in the case of $S=1$, i.e., the odd real wave function $u_{0}(x)$, see Fig. %
\ref{stableS1}(a) above. It is also worthy to stress that, while $u_{0}(x)$
virtually vanishes in a broad core area in Fig. \ref{unstableS2}(a), $%
u_{0}(x)$ keeps small positive values at all $x\neq 0$, i.e., it does not
cross zero.

Similar to what is demonstrated above for $S=1$, Fig. \ref{charS2}(a) shows
that $\beta $ gradually decreases with the increase of $C$, while $%
\varepsilon (C)$ may strongly deviate from $\varepsilon =1$, indicating
strong anisotropy of vortex solitons with $S=2$ [see in Fig. \ref{charS2}%
(b)]. The branches shown in Fig. \ref{ICOCexampleS2} are completely stable,
whereas their extensions, not shown in the figure, carry solitons which are
unstable against splitting in the discrete direction. Typical examples of
the evolution of the stable and unstable vortex solitons with $S=2$ are
displayed in Figs. \ref{unstableS2}(b,c). A comparison of values of the
Hamiltonian between the vortex solitons with $S=1$ and $2$, which have equal
powers, $P$, are shown in Figs. \ref{charS2}(c,d). The comparison indicates
that, for the same $N$, the Hamiltonian is smaller for $S=1$. Furthermore,
the existence region for the vortices with $S=1$ is larger (in some cases,
much larger) than that for their counterparts with $S=2$.

Interestingly, two different stable branches are found for $N=8$ with
smaller and larger values of $C$, which are shown by the two red-color
segments in Fig. \ref{charS2}(c). This result implies that there are two
types of vortex soliton with $S=2$ for $N=8$. To illustrate this conclusion,
Fig. \ref{ICOCexampleS2}(b) shows an example of the vortex state belonging
to the branch with larger $C$ and $N_{\mathrm{core}}=2$. An example of the
stable vortex soliton from the branch with smaller $C$ is displayed in Fig. %
\ref{unstableS2}(d), with $N_{\mathrm{core}}=4$.

\begin{figure}[h]
{\includegraphics[width=0.9\columnwidth]{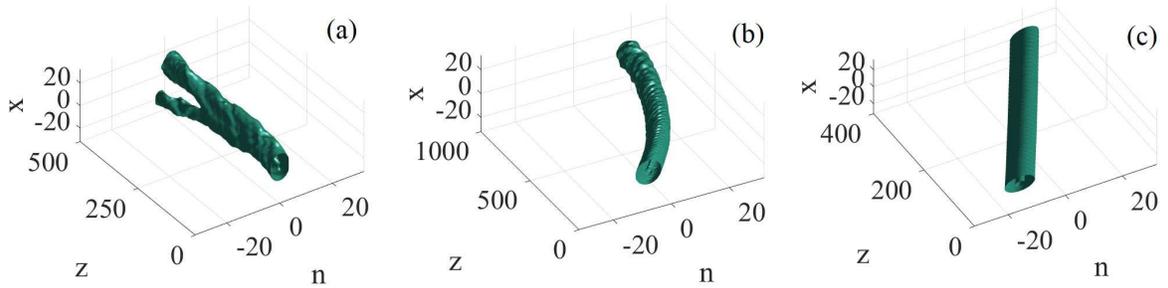}}
\caption{(a) The vortex soliton with $\left( P,C\right) =\left(
100,0.1\right) $, destroyed by the application of the kick, as per Eq. (%
\protect\ref{kick}). (b) A finite jump of the kicked vortex soliton, with $%
(P,C)=(100,0.18)$. (c) An example of a mobile vortex soliton, with $%
(P,C)=(100,0.3)$. The strength of the kick in all cases is $\protect\eta =0.1%
\protect\pi $. }
\label{moving}
\end{figure}

\begin{figure}[h]
{\includegraphics[width=0.6\columnwidth]{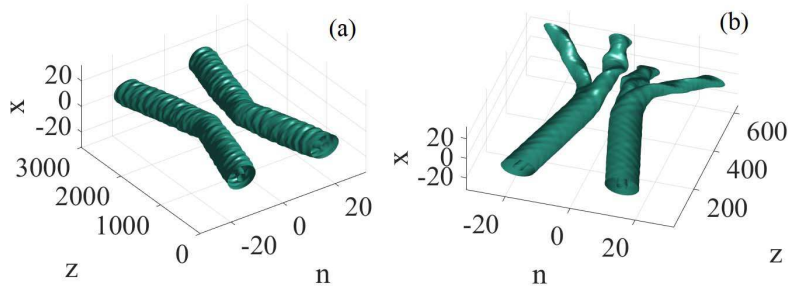}}
\caption{(a) Elastic collision between two moving semidiscrete vortex
solitons, kicked by $\protect\eta =\pm 0.01\protect\pi $. (b) Inelastic
collision between moving solitons, kicked by $\protect\eta =\pm 0.03\protect%
\pi $. In the latter case, each vortex soliton splits in two fragments after
the collision. Here, we select the vortex solitons with parameters $\left(
P,C\right) =\left( 100,0.3\right) $, originally placed at positions $%
n_{0}=\pm 16$.}
\label{moving2}
\end{figure}

\subsection{Mobility of the semidiscrete vortex solitons}

A nontrivial issue is mobility of the vortex solitons in the discrete
direction, initialized by the application of a kick, with strength $\eta $,
to them:
\begin{equation}
U_{n}(x,z=0)=U_{n}^{(0)}(x)e^{i\eta n},  \label{kick}
\end{equation}%
where $U_{n}^{(0)}(x)$ represents a quiescent soliton. Here, we address this
issue for $U_{n}^{(0)}(x)$ taken as the stable vortex of the OC type with $%
S=1$ and $P=100$. At $C<0.2$, the kick cannot set the soliton in progressive
motion, just destroying it if $\eta $ is too large, as shown in Fig. \ref%
{moving}(a), or causing a finite leap, as shown in Fig. \ref{moving}(b). The
kicked vortex soliton demonstrates mobility at moderate discreteness, with $%
C>0.2$, see an example in Fig. \ref{moving}(c).

Collisions between two vortex solitons moving in the opposite discrete
directions can be initialized by taking
\begin{equation}
U_{n}(x,z=0)=U_{+n_{0}}^{(0)}(x)e^{i\eta n}+U_{-n_{0}}^{(0)}(x)e^{-i\eta n},
\end{equation}%
where $2n_{0}$ is the initial separation between the solitons. Elastic
collisions occur if $\eta $ is small. In this case, the two colliding vortex
solitons retain their vorticities after the collision. Inelastic collisions
occur between the vortex solitons kicked with larger values of $\eta $. In
this case, the collision splits each soliton in fragments. Figure \ref%
{moving2} shows typical examples of elastic and inelastic collisions for the
vortex solitons with $(P,C,\pm n_{0})=(100,0.3,\pm 16)$. In this case, the
collision remains elastic at $\eta <0.016\pi $.

\section{Conclusion}

We have introduced the spatial-domain model for stacked set of
tunnel-coupled planar waveguides with the combination of intrinsic
self-focusing and defocusing cubic and quintic nonlinearities. The model
applies as well to arrays of tunnel-coupled fiber waveguides with the same
nonlinearity and anomalous group-velocity dispersion, which is a
temporal-domain counterpart of the paraxial diffraction in planar guiding
cores. Unlike fundamental semidiscrete solitons, previously studied in a
similar model with the cubic-only nonlinearity, we here aimed to construct
solitons with embedded vorticities, $S=1$ and $2$. It is found that such
vortex solitons of the IC and OC (intersite- and onsite-centered) types,
composed of $N$ excited sites carrying non-negligible amplitudes in the
transverse direction, form stable families, starting from minimum values $%
\left( N_{\mathrm{IC}}\right) _{\min }=2$ and $\left( N_{\mathrm{OC}}\right)
_{\min }=3$, respectively. The semidiscrete vortex solitons of the OC type
always feature an internal core, while the IC solitons with small $N$ may
have a coreless structure. The system admits multistability, i.e.,
coexistence of stable vortex solitons with equal norms and different values
of $N$. The existence and stability of such modes are only possible when the
quintic defocusing term is not too small. The vortex solitons feature
mobility in the discrete direction, provided that the coefficient of the
intersite coupling exceeds a certain minimum value. Semidiscrete vortex
solitons moving in opposite directions collide elastically or inelastically
if they are set in motion by relatively weak or strong kicks, respectively.


\begin{acknowledgments}
This work was supported by the Key Research Projects of General Colleges in
Guangdong Province through grant No. 2019KZDXM001, NNSFC (China) through
grant Nos. 11905032, 11874112, and 11575063, the Foundation for
Distinguished Young Talents in Higher Education of Guangdong through grant
No. 2018KQNCX279. The work of B.A.M. is supported, in a part, by Israel
Science Foundation through grant No. 1286/17.
\end{acknowledgments}

\end{document}